\documentclass{ws-mpla}

\def\arcsec{\hbox{$^{\prime\prime}$}}

\def\farcm{\hbox{$.\mkern-4mu^\prime$}}

\newcommand{\simgt}{\lower.5ex\hbox{$\; \buildrel > \over \sim \;$}}
\newcommand{\simlt}{\lower.5ex\hbox{$\; \buildrel < \over \sim \;$}}

\newcommand{\baredth}{\;\overline{\raise1.0pt\hbox{$'$}\hskip-6pt \partial}\;}
\newcommand{\edth}{\;\raise1.0pt\hbox{$'$}\hskip-6pt\partial\;}

\begin{document}

\markboth{K. Umetsu, M. Takada, \& T. Broadhurst}
{Probing the Cluster Mass Distribution using Subaru Weak Lensing Data}

\catchline{}{}{}{}{}

\title{
PROBING THE CLUSTER MASS DISTRIBUTION
USING SUBARU WEAK LENSING DATA
}

\author
{\footnotesize KEIICHI UMETSU$^1$, MASAHIRO TAKADA$^2$, 
TOM BROADHURST$^3$}

\address
{1. Institute of Astronomy and Astrophysics, Academia Sinica, Taipei, Taiwan\\
 2. Astronomical Institute, Tohoku University, Sendai, Japan\\
 3. School of Physics and Astronomy, Tel Aviv University, Tel Aviv,
 Israel\\
keiichi@asiaa.sinica.edu.tw}

\maketitle

\pub{Received (Day Month Year)}{Revised (Day Month Year)}

\begin{abstract}
We present results from a 
weak lensing analysis of the galaxy cluster A1689
 ($z=0.183$)
based on deep wide-field imaging data taken 
with Suprime-Cam on Subaru telescope. 
A maximum entropy method has been used to reconstruct directly
the projected mass distribution of A1689
from combined lensing distortion and magnification measurements
of red background galaxies.
The resulting mass distribution is clearly concentrated around the cD
 galaxy, and mass and light in the cluster are similarly distributed in
 terms of shape and orientation. The azimuthally-averaged mass profile
 from the two-dimensional reconstruction is in good agreement with the
 earlier results from the Subaru one-dimensional analysis of the weak
 lensing data, supporting the assumption of quasi-circular symmetry in
 the projected mass distribution of the cluster.



\end{abstract}

\ccode{PACS Nos.: include PACS Nos.}

\section{Introduction}

Weak gravitational lensing of background galaxies 
provides a unique, direct way to study the mass distribution of galaxy
clusters.\cite{BS01,UTF99}
Recent improvements in the quality of observational data usable for
lensing studies now allow an accurate determination of the mass
distribution in clusters.
A1689
is one of the best studied lensing clusters,\cite{Tom05,BTU05,OTUB,Elinor} 
located at a moderately low
redshift of $z=0.183$.
Deep HST/ACS imaging of the central region of A1689
has revealed
106 multiply lensed images of 30 background
galaxies, which allowed a detailed reconstruction of the mass
distribution in the cluster core 
($10h^{-1} {\rm kpc} \simlt r \simlt 200 h^{-1}{\rm kpc}$).\cite{Tom05}
In Ref.~\refcite{BTU05},
we developed a model-independent method for
reconstructing the cluster mass profile using azimuthally-averaged
weak-lensing distortion and magnification measurements,
and derived a projected mass profile of 
A1689 out to the cluster virial radius ($r\simlt 2 h^{-1}$ Mpc)
based on the wide-field, deep imaging data taken with 
{\it Suprime-Cam} on the 8.2m
{\it Subaru telescope}.  
The combined strong and weak lensing mass profile is well fitted by an
NFW\cite{NFW} profile with high concentration of $c_{\rm vir}\sim 13.7$,
which is significantly larger than theoretically expected
($c_{\rm vir}\simeq 4$) for the standard LCDM model.\cite{Bullock}

In this paper 
we present a weak lensing analysis of 
A1689 using wide-field Subaru imaging data,
with special attention to the map-making process.
Throughout this paper,
we use the AB magnitude system, and
adopt
the concordance $\Lambda$CDM cosmology with
($\Omega_{\rm m0}=0.3$, $\Omega_{\lambda 0}=0.7$, $h=0.7$). 
In this cosmology one arcminute corresponds to the physical scale
$129$kpc$/h$ for this cluster.

\section{Sample selection} 
\label{data}

For our weak lensing analysis
we used Subaru/Suprime-Cam imaging data of A1689 in $V$ (1,920s) 
and SDSS $i'$ (2,640s) 
retrieved from the Subaru archive,
SMOKA (see Ref.~\refcite{BTU05}, \refcite{Elinor} for more details).
The FWHM in the final co-added image is 
$0\arcsec\!\!.82$ in $V$ and 
$0\arcsec\!\!.88$ in $i'$ with $0\arcsec\!\!.202$
pix$^{-1}$, covering a field of $30'\times 25'$.
The limiting magnitudes are
$V=26.5$ and $i'=25.9$ for a $3\sigma$ detection within a $2$\arcsec
aperture.
A careful background selection is critical 
for a weak lensing analysis.\cite{BTU05,Elinor}
For the number counts to measure magnification, we define a sample of 
8,907 galaxies ($\bar{n}_{\mu}=12.0$ arcmin$^{-2}$) with $V-i'>1.0$. 
For distortion measurement, 
we define a sample of 5,729 galaxies ($\bar{n}_{g}=7.59$ arcmin$^{-2}$)
with colors $0.22$ mag redder than the color-magnitude sequence of
cluster E/S0 galaxies, $(V-i') +0.0209i'-1.255 > 0.22$. The smaller
sample is due to the fact that distortion analysis requires galaxies
used are well resolved to make reliable shape measurement. 
We adopt a limit of $i'<25.5$  to avoid incompleteness effect. 
In what follows we will assume $\langle z_s\rangle = 1$ for the mean
redshift of the red galaxies,\cite{Tom05,BTU05}
but note that the low redshift of 
A1689 means that for
lensing work, a precise knowledge of this redshift is not critical.


\begin{figure}[h]
 \begin{center}
  \includegraphics[width=70mm, angle=270]{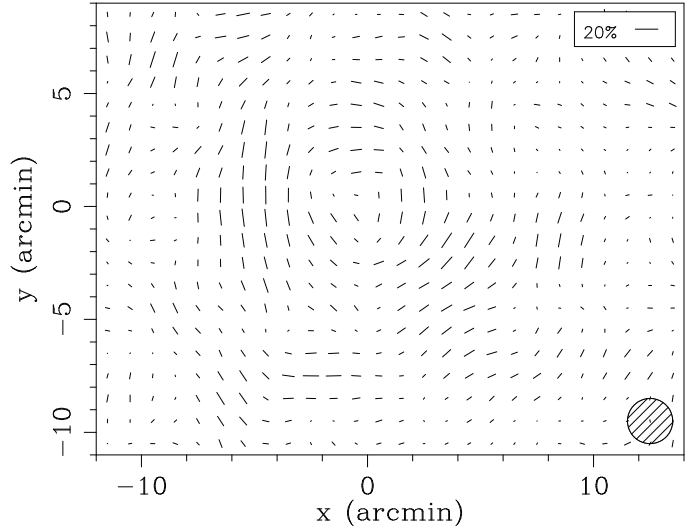}
 \end{center}
%
\caption{Gravitational reduced-shear field in A1689 obtained from shape
 distortions of the red background galaxies, smoothed with a
 Gaussian with ${\rm FWHM} = 2'$ for visualization purposes. 
The shaded circle indicates the FWHM of the Gaussian.
The coordinate origin is at the optical center (see Figure
 \ref{fig:image}). 
}
\label{fig:shear}
\end{figure}

\section{Lensing Distortions}
\label{shear}

We use the IMCAT package developed by N. Kaiser
\footnote{http://www.ifa.hawaii/kaiser/IMCAT} to perform object
detection, photometry and shape measurements, following the formalism
outlined in  Ref.~\refcite{KSB}.
We have modified the method somewhat following the procedures
described in Ref.~\refcite{Erben}.
To obtain an estimate of the reduced shear,
$g_{\alpha}=\gamma_{\alpha}/(1-\kappa)$, 
we measure the image ellipticity $e_{\alpha}$
from the weighted quadrupole moments of the surface brightness of
individual galaxies. 
Firstly the PSF anisotropy needs to be corrected using the star images
as references:
\begin{equation}
e'_{\alpha} = e_{\alpha} - P_{sm}^{\alpha \beta} q^*_{\beta}
\label{eq:qstar}
\end{equation}
where $P_{sm}$ is the {\it smear polarizability} tensor  
being close to diagonal, and
$q^*_{\alpha} = (P_{*sm})^{-1}_{\alpha \beta}e_*^{\beta}$ 
is the stellar anisotropy kernel.
We select bright, unsaturated foreground stars identified in a branch
of the half-light radius ($r_h$) vs. magnitude ($i'$) diagram
($20<i'<22.5$, $\left<r_h\right>_{\rm median}=2.38$ pixels) to
calculate $q^*_{\alpha}$.

In order to obtain a smooth map of $q^*_{\alpha}$ which is used in
equation (\ref{eq:qstar}), we divided the $9{\rm K}\times 7.4{\rm K}$
image into $5\times 4$ chunks each with $1.8{\rm K}\times 1.85{\rm K}$
pixels, and then fitted the $q^*$ in each chunk independently with
second-order bi-polynomials, $q_*^{\alpha}(\vec{\theta})$, in
conjunction with iterative $\sigma$-clipping rejection on each
component of the residual
$e^*_{\alpha}-P_{*sm}^{\alpha\beta}q^*_{\beta}(\vec{\theta})$.  The
final stellar sample consists of 540 stars, or the mean surface number
density of $\bar{n}_*=0.72$ arcmin$^{-2}$. From the rest of the object
catalog, we select objects with $2.4 \lesssim r_h \lesssim 15$ pixels
as an $i'$-selected weak lensing galaxy sample, which contains $61,115$
galaxies or $\bar{n}_g\simeq 81$ arcmin$^{-2}$.
It is worth
noting that the mean stellar ellipticity before correction is
$(\bar{e_1}^*, \bar{e_2}^*) \simeq (-0.013, -0.018)$ 
over the data field, while the residual
$e^*_{\alpha}$ after correction
is reduced to $ {\bar{e}^{*{\rm res}}_1} = (0.47\pm
1.32)\times 10^{-4}$, $ {\bar{e}^{*{\rm res}}_2} = (0.54\pm
0.94)\times 10^{-4}$.
The mean offset from the null expectation is 
$|\bar{e}^{* \rm res}| = (0.71\pm 1.12) \times 10^{-4}$.
On the other hand, the rms value of stellar ellipticities,
$\sigma_{e*}\equiv\left<|e^*|^2\right>$, is reduced from $2.64\%$ to
$0.38\%$ when applying the anisotropic PSF correction.
Second, we need to correct the isotropic smearing effect on 
image ellipticities
caused by seeing and the window function used for the shape
measurements. The pre-seeing reduced shear $g_\alpha$ can be
estimated from 
\begin{equation}
\label{eq:raw_g}
g_{\alpha} =(P_g^{-1})_{\alpha\beta} e'_{\beta}
\end{equation}
with the {\it pre-seeing shear polarizability} tensor
$P^g_{\alpha\beta}$.
We follow the procedure described in Ref.~\refcite{Erben} to measure
$P^g$.
We adopt the scalar correction scheme,\cite{Erben,Hoekstra98,Hudson98}
namely,
$P^g_{\alpha\beta}=\frac{1}{2}{\rm tr}[P^g]\delta_{\alpha\beta}\equiv
P^g_{\rm s}\delta_{\alpha\beta}$.
The $P_{g}^{\rm s}$ measured for individual objects are still noisy
especially for small and faint objects.  
We remove from the galaxy catalog those objects
that yield a negative value of $P_g^{\rm s}$
estimate to avoid noisy shear estimates.
We then adopt a smoothing
scheme in object parameter space.\cite{Erben,vanWaerbeke00,Hamana03}
We first identify thirty neighbors for each object in
$r_g$-$i'$ parameter space.
We then calculate over the local ensemble
the median value $\langle P_g^{\rm s}\rangle$
of $P_g^{\rm s}$ 
and the variance $\sigma^2_{g}$ 
of $g=g_1+ig_2$ using equation (\ref{eq:raw_g}).
The dispersion $\sigma_g$ is used as an rms error of the shear estimate
for individual galaxies.
The mean variance $\bar{\sigma}_g^2$ over the red galaxy sample is obtained as
$\simeq 0.133$, or $\sqrt{\bar{\sigma}_g^2}\approx 0.36$. 
Finally, we use the following estimator for 
the reduced shear:
$g_{\alpha} = e'_{\alpha}/\left< P_g^{\rm s}\right>$.

For map-making,
we then pixelize the distortion data into a regular grid
of $N_{\rm pix}=21\times 17$ independent pixels, 
covering a field of $\approx 30'\times 24'$.
The pixel size is $\Delta_{\rm pix}=1\farcm 4$, and the mean
galaxy counts per pixel is $\sim 15$.
The bin-averaged reduced shear is given as 
$\bar{g}_{\alpha,i}\equiv 
\bar{g}_{\alpha}(\vec{\theta}_i)=\sum_{k\in {\rm cell} i}
u_k g_{\alpha,k}/\sum_{k\in {\rm cell} i} u_k$ $(a=1,2; i=1,2,...,N_{\rm pix})$,
where 
$g_{\alpha,k}$ is the estimate of the $\alpha$th component of the
reduced shear for the $k$th galaxy, and 
$u_k=1/(\sigma_{g,k}^2+\alpha^2)$ is its inverse-variance weight
softened with a constant $\alpha$. Here we choose
$\alpha=0.4$.\cite{Hamana03}
In Figure \ref{fig:shear} we show the reduced-shear field 
obtained from the red galaxy sample, where for visualization purposes
the $\bar{g}_{\alpha}(\vec{\theta})$ is resampled on to a finer grid and
smoothed with a Gaussian with ${\rm FWHM}=2'$.

\begin{figure}[htbp]
 \begin{center}
   \includegraphics[width=70mm, angle=270]{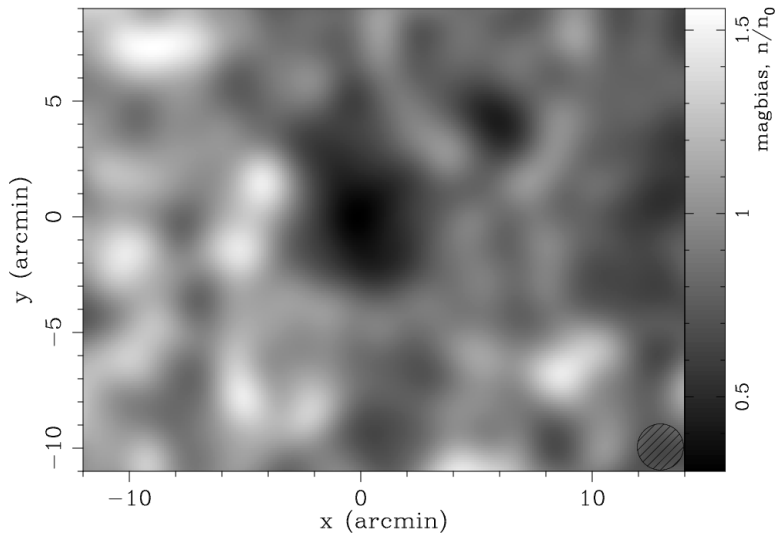}
 \end{center}
\caption{Distribution of the lensing magnification bias 
$n/n_0$
measured from a red-galaxy sample in the background of A1689, 
smoothed with a Gaussian 
 with ${\rm FWHM} = 2'$ for visualization purposes.
The shaded circle indicates the FWHM of the Gaussian.}
\label{fig:magbias}
\end{figure}

\section{Magnification Bias}\label{mag}

Lensing magnification, $\mu(\vec\theta)$, influences the observed
surface density of background galaxies, expanding the area of
sky, and enhancing the flux of galaxies.
In the sub-critical regime,\cite{BS01,UTF99}
the magnification $\mu$ is given by 
$\mu=1/\left[(1-\kappa)^2-\gamma_1^2-\gamma_2^2\right]$.
The count-in-cell statistics are measured from 
the flux-limited red galaxy sample 
(see \S \ref{data}) on the same grid as the distortion data:
 $N_i\equiv N(<m_{\rm cut}; \vec{\theta}_i)=\sum_{k\in {\rm cell} i}1$
with $m_{\rm cut}$ being the magnitude cutoff corresponding to the
flux-limit.
The normalization and slope of the unlensed number counts
$N_0(<m_{\rm cut})$ for our red galaxy sample are reliably estimated as 
$n_{\mu,0}=12.6\pm 0.23$ arcmin$^{-2}$ and 
$s\equiv d\log{N_0(m)}/dm=0.22\pm 0.03$
from the outer region $\ge 10'$.\cite{BTU05}
The slope is less than the lensing invariant slope, $s=0.4$, 
and hence a net deficit of background galaxies is expected:
$N_i/N_0=n_{\mu,i}/n_{\mu,0}=\mu_i^{2.5s-1}$.
The masking effect due to bright cluster galaxies is properly taken into
account and corrected for.\cite{BTU05}
Figure \ref{fig:magbias} shows a clear depletion of the red galaxy
counts in the central, high-density region of the cluster.
Note we have ignored the intrinsic clustering of
background galaxies, which seems a good approximation,\cite{BTU05}
though some
variance is apparent in the spatial distribution of red galaxies.

\section{Two-Dimensional Mass Reconstruction}\label{massrec}

\begin{figure}[h]
 \begin{center}
   \includegraphics[width=90mm]{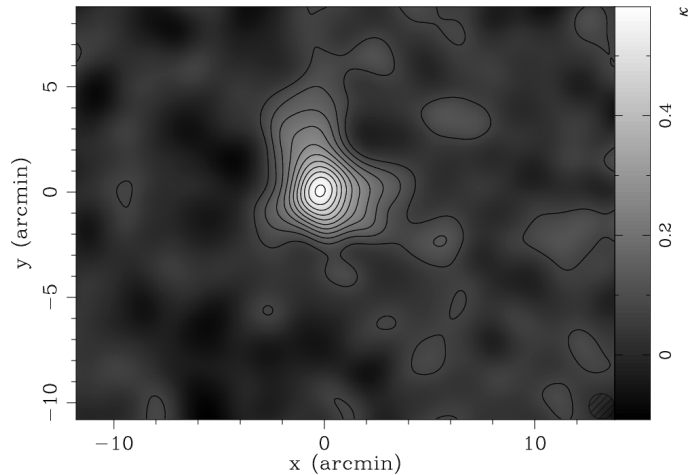}
 \end{center} 
\caption{
The projected mass distribution $\kappa$ of A1689 reconstructed 
by a maximum entropy method
using weak-lensing distortion and
 magnification data of red background galaxies. 
For visualization purposes the $\kappa$-map is
resampled on to a finer grid, and 
 smoothed with a Gaussian with ${\rm FWHM}=2'$. The lowest contour is
 $\kappa=0.07$, and the contour steps are $\Delta \kappa=0.05$.}
\label{fig:2dmass}
\end{figure}

\begin{figure}[ht] 
 \begin{center}
   \includegraphics[width=70mm]{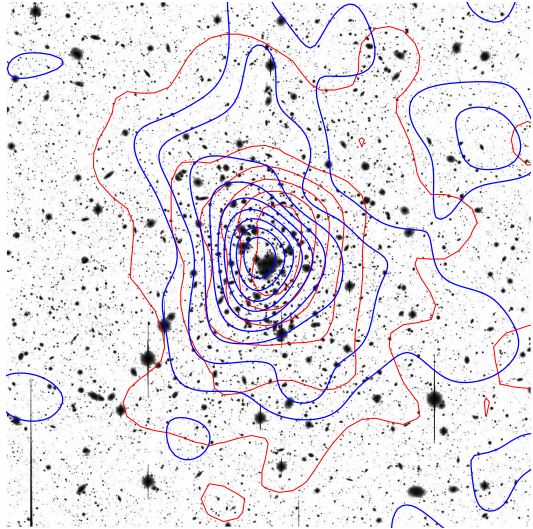}
 \end{center} 
\caption{Contours ({\it thick}) of the reconstructed 
lensing convergence $\kappa$ 
superposed on the
$i'$-band image of the central region of A1689. 
The field size is $15'$ on a side.
The lowest contour
 and the contour intervals are $0.05$.
Also superposed are the contours ({\it thin}) of the projected
 luminosity density of the cluster member galaxies defined by the
 color-magnitude sequence. North is to the top and East is to the left.}
\label{fig:image}
\end{figure}

The relation between distortion and convergence is non-local, and
the convergence $\kappa$
derived from distortion data alone suffers from a mass sheet
degeneracy.\cite{BS01,UTF99} 
However, by combining the distortion and magnification
measurements the convergence can be obtained unambiguously with the
correct mass normalization. 
Here we combine pixelized distortion and magnification data of the red
background galaxies,
and reconstruct the two-dimensional (2D) 
distribution of $\kappa$  using a maximum
entropy method (MEM) extended to account for positive/negative
distributions of the underlying field.\cite{MEM,MEM_HL98}
We take into account the non-linear, but sub-critical, regime of the
lensing properties, $\kappa$ and $\gamma_{\alpha}$.
We take $\kappa_i=\kappa(\vec{\theta}_i)$ 
as the {\it image} to be reconstructed,\cite{MEM}
and express a set of discretized $\kappa$-values as ${\bf
p}=\{\kappa_i\}$ $(i=1,2,...,N_{\rm pix})$.
The total log-likelihood function, $F({\bf p})=-\ln{{\cal L}({\bf p})}$, 
is expressed as a 
linear sum of the shear/magnification data log-likelihoods\cite{Schneider00} 
and the entropy term\cite{MEM}:
\begin{eqnarray}
\label{eq:loglikelihood}
F({\bf p})& = &l_g({\bf p}) + l_{\mu}({\bf p}) - \alpha S({\bf p}, {\bf
m}),\\
l_g &\equiv& -\ln{{\cal L}_g}
\approx \sum_{i=1}^{N_{\rm pix}}
\sum_{\alpha=1}^{2} 
\frac{ (\bar{g}_{\alpha,i}-\hat{g}_{\alpha,i}({\bf p}) )^2 } {
\sigma_{g,i}^2 },\\
l_{\mu} &\equiv& -\ln{{\cal L}_{\mu}}
\approx \frac{1}{2}\sum_{i=1}^{N_{\rm pix}}
\frac{(N_{i}-\hat{N}_{i}({\bf p}))^2} {N_i},
\end{eqnarray}
where $\hat{g}_{\alpha,i}({\bf p})$ and  $\hat{N}_i({\bf p})$
are the theoretical expectations for 
$\bar{g}_{\alpha,i}$ and $N_i$, respectively, 
$\sigma_{g,i}\equiv \sigma_g(\vec{\theta}_i)$ is the rms error for 
$\bar{g}_{i}=\bar{g}_{1}(\vec{\theta}_i)+i\bar{g}_{2}(\vec{\theta}_i)$,
and $S({\bf p}, {\bf m})$ is the cross entropy function 
for the positive/negative field;\cite{MEM}
the ${\bf m}$ is a set of the model parameters and $\alpha (>0)$ is the
regularization constant. 
The maximum likelihood solution, $\hat{\bf p}$,
is obtained by minimizing
the function $F({\bf p})$ with respect to ${\bf p}$
for given $\alpha$ and ${\bf m}$. 
We take $m_i={\rm const}\equiv m$, and determine by iteration
the Bayesian value of $\alpha$ for a given value of  $m$. 
We found that 
the maximum-likelihood solution for the Bayesian $\alpha$ 
is insensitive to the choice of $m$. In the following we set $m$ to be
$0.2$. We note that the adopted MEM prior\cite{MEM,MEM_HL98} ensures 
$\kappa_i \to 0$ in the noise-dominated regime, $F({\bf p})\sim -\alpha
S({\bf p},{\bf m})$
(i.e., maximizing the entropy alone).
In order to quantify the errors on the mass reconstruction we
evaluate the Hessian matrix of the function $F({\bf p})$ at ${\bf
p}=\hat{\bf p}$,
$H_{ij}(\hat{\bf p})=\frac{\partial^2 F({\bf p})}
{\partial p_i \partial p_j}|_{{\bf p}=\hat{\bf p}}$, 
from which the covariance matrix of the
parameters ${\bf p}$ is given by $C_{ij}\equiv \langle \Delta\kappa_i
\Delta\kappa_j \rangle=
(H)^{-1}_{ij}(\hat{\bf p})$.
Figure \ref{fig:2dmass} displays the $\kappa$ map reconstructed
with the MEM method.
In Figure \ref{fig:image} we compare the contours of the reconstructed
$\kappa$ ({\it thick})
and the $i'$-band luminosity density of the cluster sequence galaxies
({\it thin}) superposed on the
$i'$-band image of the central region of A1689.


\section{Model-Independent Mass Profile of A1689}\label{mass}

\begin{figure}[h]
 \begin{center}
  \includegraphics[width=70mm, angle=270]{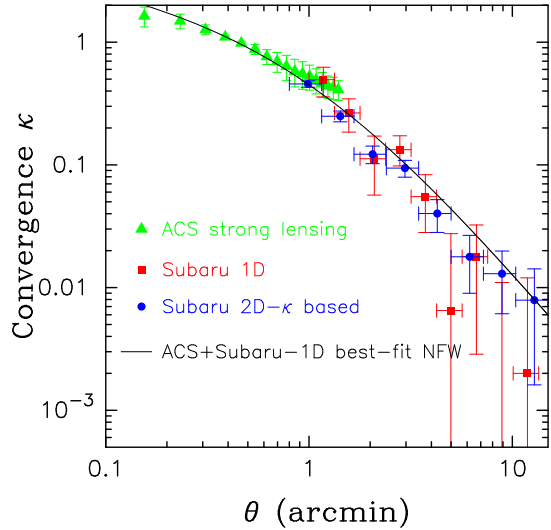}
 \end{center} 
\caption{
Reconstructed mass profile of A1689.
The filled circles with error bars represent the results 
based on the 2D $\kappa$ map reconstructed with the MEM algorithm.
The error bars are correlated.
The triangle and square symbols
with error bars show the results from the HST/ACS strong lensing
analysis (Broadhurst et al. 2005, ApJ, 621, 53)
and 
the Subaru weak lensing analysis 
with the 1D reconstruction method
(Broadhurst, Takada, Umetsu et
 al. 2005, ApJ, 619, L143, hereafter BTU05)
respectively. 
The solid curve shows 
the best-fitting NFW profile for the combined ACS+Subaru data
by BTU05 ($M_{\rm vir}=1.93\times 10^{15} M_{\odot}, c_{\rm
 vir}=13.7$).
The
 best-fitting NFW profile for the ACS+Subaru profile
 has a high concentration, $c_{\rm vir}=13.7$,
 and somewhat overestimates the inner slope and is a bit
 shallower than the 1D-based results at large radius. 
}
\label{fig:mass}
\end{figure}

We show in Figure \ref{fig:mass} the azimuthally-averaged profile 
of the reconstructed convergence, $\kappa(\theta)$,
as a function of projected radius $\theta$  
from the optical center of A1689 ({\it circles}).
The vertical error bars represent the $1\sigma$ uncertainties based on
the error covariance matrix $C_{ij}$ of the reconstruction.
Note that the error bars are correlated.
Also shown for comparison are the earlier results 
from the HST/ACS strong lensing analysis 
({\it triangles})
and 
the Subaru weak lensing analysis ({\it squares})
with the one-dimensional (1D) reconstruction method,
respectively, along with the best-fitting NFW\cite{NFW} model
({\it solid})
for the combined ACS+Subaru profile
(see Ref.~\refcite{BTU05}).

\section{Discussion and Conclusions}\label{con}

We presented results from our weak lensing analysis of A1689 based on
deep wide-field imaging data taken with Subaru/Suprime-Cam.
We used a MEM algorithm to reconstruct the projected mass map in A1689
from combined distortion and magnification data of our red background
galaxy sample.
The combination of distortion and
 magnification data breaks the mass sheet degeneracy inherent in all
 reconstruction methods based on distortion information alone.
Our results show that 
mass and light in A1689
are similarly distributed in terms of shape and orientation, 
and clearly concentrated around the cD galaxy (see Figure \ref{fig:image}).
The resulting mass profile from the present full 2D 
reconstruction 
is in good agreement with the 
results from the earlier Subaru 1D
analysis\cite{BTU05} (see Figure \ref{fig:mass}),
supporting the assumption of
quasi-circular symmetry in the projected mass distribution.




\section*{Acknowledgments}

Part of this work is based on  
data collected at the Subaru Telescope,
which is operated by the National Astronomical Society of Japan.
The work is in part supported by the National Science Council of Taiwan
under the grant NSC95-2112-M-001-074-MY2.

\end{document}